\documentclass[preprint2,times,tighten]{aastex}

\usepackage{amsmath,amstext}
\usepackage[figure,figure*]{hypcap}
\usepackage{xspace}
\usepackage{newtxmath} 

\newcommand{\etc}{\textit{etc}\xspace}
\newcommand{\IRSSquare}{IRS$^2$\xspace}
\newcommand{\IRSSquarePars}[2]{$\left(n= #1 ,r= #2 \right)$}
\newcommand{\JWST}{\textit{JWST}\xspace}

\newcommand{\eg}{\textit{e.g.}\xspace}
\newcommand{\ie}{\textit{i.e.}\xspace}
\newcommand{\period}{}
\renewcommand{\vec}[1]{\mathbf{#1}}

\shorttitle{Improved Reference Sampling and Subtraction}
\shortauthors{Bernard J. Rauscher et al.}
\slugcomment{Accepted by Publications of the Astronomical Society of the Pacific}

\begin{document}

\title{Improved Reference Sampling and Subtraction:\\ A Technique for Reducing the Read Noise of Near-infrared Detector Systems}
\author{Bernard J. Rauscher\altaffilmark{1}, Richard G. Arendt\altaffilmark{2}, D.J. Fixsen\altaffilmark{3}, Matthew A. Greenhouse\altaffilmark{1}, Matthew Lander\altaffilmark{7}, \\Don Lindler\altaffilmark{4}, Markus Loose\altaffilmark{5}, S.H. Moseley\altaffilmark{1}, D. Brent Mott\altaffilmark{6}, Yiting Wen\altaffilmark{6}, Donna V. Wilson\altaffilmark{7}, \\and Christos Xenophontos\altaffilmark{7}}
\affil{NASA Goddard Space Flight Center, 8800 Greenbelt Road, Greenbelt, MD 20771, USA}
\altaffiltext{1}{NASA/GSFC, Observational Cosmology Laboratory, Code 665}
\altaffiltext{2}{CRESST/UMBC/GSFC,Greenbelt, MD}
\altaffiltext{3}{CRESST/UMd/GSFC, Greenbelt, MD}
\altaffiltext{4}{Sigma Space Corporation/GSFC, Greenbelt, MD}
\altaffiltext{5}{Markury Scientific, Inc., 518 Oakhampton Street, Thousand Oaks, CA}
\altaffiltext{6}{NASA/GSFC, Detector Systems Branch, Code 553}
\altaffiltext{7}{NASA/GSFC, Flight Software Systems Branch, Code 582}
\email{Bernard.J.Rauscher@nasa.gov}

\begin{abstract}
Near-infrared array detectors, like the \JWST NIRSpec's Teledyne's H2RGs,  often provide reference pixels and a reference output. These are used to remove correlated noise. Improved Reference Sampling and Subtraction (\IRSSquare; pronounced ``IRS-square'') is a statistical technique for using this reference information optimally in a least squares sense. Compared to  ``traditional'' H2RG readout, \IRSSquare uses a different clocking pattern to interleave many more reference pixels into the data than is otherwise possible. Compared to standard reference correction techniques, \IRSSquare subtracts the reference pixels and reference output using a statistically optimized set of frequency dependent weights. The benefits include somewhat lower noise variance and much less obvious correlated noise. NIRSpec's \IRSSquare images are cosmetically clean, with less $1/f$ banding than in traditional data from the same system. This article describes the \IRSSquare clocking pattern and presents the equations that are needed to use \IRSSquare in systems  other than NIRSpec. For NIRSpec, applying these equations is already an option in the calibration pipeline. As an aid to instrument builders, we provide our prototype \IRSSquare calibration software and sample \JWST NIRSpec data. The same techniques are applicable to other detector systems, including those based on Teledyne's H4RG arrays. The H4RG's ``interleaved reference pixel readout'' mode is effectively one \IRSSquare pattern.
\end{abstract}


\section{Introduction}

The Near Infrared Spectrograph \cite[NIRSpec;][]{Birkmann:2016ck} is the James Webb Space Telescope's (\JWST) primary $0.6-5~\mu$m spectrograph. NIRSpec's main mode is multi-object spectroscopy with spectral resolution $R=\lambda/\Delta\lambda = 100$, 1000, and 2700. Due to the low background  provided by the observatory and the low dark current rates of the detectors, NIRSpec will be detector noise limited for most faint object observations. In this regime, the exposure time needed to achieve a given signal-to-noise ratio for a given pixel scales directly with the read noise. Minimizing read noise is therefore key to maximizing NIRSpec's performance.

Recovering the spectrum of faint objects usually involves operations on more than one pixel. We therefore identified correlated noise ($1/f$ banding, \etc) as a second noise feature that  limits NIRSpec's sensitivity. Correlated noise is particularly important for NIRSpec's multi-object spectrograph (MOS) mode because having less correlated noise enables more efficient spectral extraction. In particular, it reduces the need for local sky samples to mitigate correlated noise.

If useful local sky is always available (\eg in a sparse Deep Field), then one can can achieve sensitivity within a few percent of \IRSSquare in NIRSpec's traditional mode. However, this places requirements on the astronomical scene. Achieving lower correlated noise has the potential to increase NIRSpec's MOS multiplex advantage in more complex scenes by making non-local sky samples more competitive.

In this context, we developed Improved Reference Sampling and Subtraction (\IRSSquare; pronounced ``IRS-square'') to reduce the read noise of NIRSpec's Teledyne SIDECAR$^\textrm{TM}$ ASIC (hereafter ``SIDECAR'') and H2RG based detector system to below what is possible in \JWST's traditional ``MULTIACCUM'' readout \citep{Rauscher:2007gc}. \IRSSquare works by making more efficient use of the H2RG's reference pixels and reference output than is possible on conventional readout schemes such as MULTIACCUM. Although we developed \IRSSquare for NIRSpec's system, it is applicable to other detector types and controllers.\footnote{Better detectors and better controllers are always beneficial. We anticipate that one could do even better than was achieved in NIRSpec by using lower noise controllers and detectors, together with \IRSSquare to reject correlated noise irrespective of its origin in the detector or controller.}

For example, Teledyne's new H4RGs build in one \IRSSquare readout pattern. In Teledyne  documentation, this appears as ``interleaved reference pixel readout''. Taking best advantage of the new mode requires the equations that are presented in $\S$~\ref{sec-equations}. Even if a SIDECAR is not used for HxRG\footnote{We define ``HxRG'' as a generic identifier for Teledyne's H4RG, H2RG, and H1RG near-IR array detectors. Apart from pixel count, the architecture is broadly similar within the HxRG family.} control, \IRSSquare can still be implemented. If using an H2RG, one would program the IR array controller to use the H2RG's internal shift registers as is described in $\S$~\ref{sec-clocking}. We believe that a similar approach could be taken with other HxRGs and, depending on the details, perhaps IR arrays from other vendors.

\begin{figure}[]
\begin{center}
\includegraphics[width=\columnwidth]{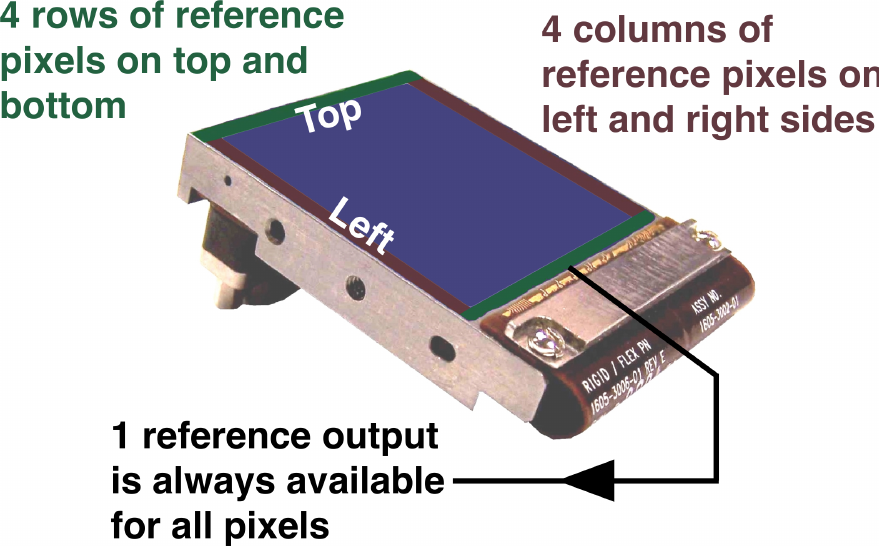}
\caption{Teledyne's H2RG detector array provides reference pixels and a reference output. Although they do not respond to light, these references are designed to electronically mimic regular pixels. As such, they can be used to remove correlated noise. There are four rows of reference pixels along the ``top'' and ``bottom'' edges, and four columns or reference pixels on the ``left'' and ``right'' sides. The reference output is physically located at one edge of the array. It has its own video output and is available at all times for all pixels. The H2RG produces $2048\times 2048$~pixel full-frame images. However, only pixels [4:2044,4:2044] (using zero-offset python notation) respond to light.}\label{fig-H2RG}
\end{center}
\end{figure}

NIRSpec uses two $2048\times 2048$~pixel Teledyne H2RG detectors. The HgCdTe detector material is sensitive over the $0.6-5~\mu$m bandpass. The H2RG (Figure~\ref{fig-H2RG}) builds in two kinds of reference pixels. Of the $2048\times 2048$~pixel image area, the outer four rows and columns of pixels on all sides are reference pixels. Although they do not respond to light, the reference pixels are designed to electronically mimic a regular light sensitive pixel. In particular, they include a dummy capacitor that simulates a regular pixel's capacitance and that is connected to the important detector substrate (DSUB) bias voltage.

Each ``frame'' of H2RG data therefore consists of the $2040\times 2040$ photosensitive ``regular pixels'' surrounded by a four pixel wide border of reference pixels on all sides. In addition to these reference pixels, there is a dedicated ``reference output'', that also mimics a regular pixel. Because the reference output is available at all times, it enables subtracting high frequency common mode noise that is missed by the reference pixels.

Teledyne H2RGs and SIDECAR ASICs are widely used at observatories around the world today. Most often, one SIDECAR is paired with each H2RG and used to read it out differentially. This is accomplished by routing the H2RG's video outputs to the SIDECAR's video inputs. The H2RG's one reference output is typically also carried back to the SIDECAR, where it is routed in SIDECAR firmware to provide the negative input to the SIDECAR's differential video preamplifiers. The  reference output is used in parallel for all $n$ video channels, where in the H2RG the number of video channels is software selectable within $n \in \{1,4,32\}$. When operated this way, the reference output is subtracted with unity gain and completely open bandwidth from all $n$ video channels within the first preamplifiers.

In contrast to the reference output, which is typically subtracted ``on the fly'' as described above (if it is used at all), the reference pixels are usually subtracted as part of the post processing to calibrate scientific data.  Although each user has their own preferred way of handling the reference pixels, there are some commonalities. Most often, each video output is treated separately from the others, and some combination of the reference pixels in rows is robustly averaged (to reject statistical outliers) and subtracted. Although only two of the video outputs have reference columns, these can also be used. When the reference columns are used, they are typically averaged and smoothed prior to applying a row-by-row reference correction to the image. Many groups arrive at the best ``recipe'' by iteratively working toward an algorithm that works well for their system.

\IRSSquare is different. The \JWST NIRSpec detector subsystem was designed to be highly linear. It was likewise designed to be extremely stable and repeatable, with the statistical properties of the noise being independent of time. These properties have been verified by test. Building on this foundation, we realized that the NIRSpec detector subsystem could be well modeled as a covariance stationary linear system. Since the reference pixels and reference outputs are designed to mimic a regular pixel except insofar as its response to light, we assumed that in the absence of light, the dark regular pixels could be represented by a linear combination of the reference output and the reference pixels. Because the system is covariance stationary to a high degree of approximation, \IRSSquare was developed in Fourier space. For a covariance stationary system, noise that appears correlated in the time domain (pixel space) is completely uncorrelated in Fourier space. 

For gauss random read noise, \IRSSquare rigorously provides the best possible reference correction if least squares is accepted as the figure of merit. This is important because it means that one knows that further reference subtraction trade studies will not provide any further benefit unless some other figure of merit is adopted. For astronomical detector systems that: (1) are linear, (2) have read noise that is statistically independent of time, and (3) have gaussian read noise after reference correction; \IRSSquare provides the best possible reference correction when least squares is used as the figure of merit. 

Even using these techniques, early trade studies revealed that the flight system had significant correlated noise extending from DC up to $\sim 3$~kHz on the low frequency side and at the 50~kHz Nyquist frequency. For this reason, we introduced a new clocking pattern to acquire many more reference samples than is possible in traditional H2RG readout. The \IRSSquare clocking pattern, which is described in $\S$~\ref{sec-clocking}, interleaves many more reference pixels than the conventional pattern. It uses these, and the reference output, to remove the $\lesssim 3$~kHz and 50~kHz noise (Figure~\ref{fig-alphas-betas}).

To summarize, the essential elements of \IRSSquare compared to traditional H2RG readout are as follows.

\begin{enumerate}
\item \IRSSquare uses a different clocking pattern to interleave many more reference pixels into the data than is otherwise possible.
\item \IRSSquare subtracts the reference pixels and reference output using an optimal set of frequency dependent weights that represent a least squares fit of these references to a training data set.
\end{enumerate}
Table~\ref{tab-comparison} briefly summarizes some of the key differences between traditional H2RG readout and \IRSSquare.

\floattable
\begin{deluxetable}{l|c|c}
\tablecaption{Comparing Traditional and \IRSSquare Readout in \JWST NIRSpec\label{tab-comparison}}
\tablehead{
\colhead{Item} & \colhead{Traditional} & \colhead{\IRSSquare}
}
\startdata
Reference pixels in rows & 
   \parbox[t][][t]{0.3\textwidth}{Four rows of reference pixels on ``top'' and ``bottom'' of every frame} &
   \parbox[t][][t]{0.3\textwidth}{Four rows of reference pixels on ``top'' and ``bottom'' of every
   frame. In addition, $r$ reference pixels from one reference row interleaved every $n$ normal pixels
   throughout the frame. For NIRSpec, the standard values are $\left(n=16,r=4\right)$. The interleaved
   reference pixels sample both even and odd numbered columns to allow removal of alternating
   column pattern noise.\vspace{3pt}} \\
Reference pixels in columns &
   \parbox[t][][t]{0.3\textwidth}{Four columns of reference pixels in ``left'' and ``right'' outputs only.\vspace{3pt}} &
   Same \\
Reference output & 
    \parbox[t][][t]{0.3\textwidth}{The reference output is subtracted in real time and with unity gain
    from each video output. It provides the reference for the SIDECAR ASIC's differential video input
    amplifiers.} &
    \parbox[t][][t]{0.3\textwidth}{The reference output is digitized in parallel with the four other video
    outputs and saved for later \IRSSquare processing. During readout, the SIDECAR ASIC's single
    ended video inputs are used. The \IRSSquare processing applies an optimal set of frequency
    dependent weights before the reference output is subtracted in post-processing.\vspace{3pt}} \\
 \enddata
\end{deluxetable}


This article is intended to provide a concise, yet reasonably complete description of \IRSSquare. It is the culmination of six years of work by a large team. As such, it reflects a mature understanding of what \IRSSquare does and why. Readers wishing to see more of the intermediate steps and background investigations may read some of our earlier papers. These include \citet{Moseley:2010kc}, which provides the first published description of the \IRSSquare concept. \citet{Rauscher:2011jx} describes early proof of concept testing using an engineering grade H2RG and SIDECAR ASIC. In \citet{Rauscher:2012cp}, we presented the full set of \IRSSquare equations in their current form for the first time. Finally, in \citet{Rauscher:2013ht} we perform principal components analysis of a flight like NIRSpec system and lay the foundations for extending \IRSSquare to remove the small amount of non-stationary noise that remains even after \IRSSquare processing.

For readers who may be deciding whether or not to try \IRSSquare, \S~\ref{sec-benefits} explains the benefits from a \JWST NIRSpec perspective. Our aim is to provide enough information (between this narrative and the freely downloadable software and sample data) for readers to make an informed decision on whether not to try \IRSSquare.

In \S~\ref{sec-concept}, we explain more of the underlying physical rationale and mathematical concepts of \IRSSquare. This section explains why \IRSSquare is expressed most naturally in Fourier space, and why we believe that the same concepts are applicable to other detector systems.

In \S~\ref{sec-clocking}, we provide an overview of the \IRSSquare implementation within the \JWST SIDECAR assembly code.   This section also explains the NIRSpec data format, which is necessary to understanding the sample data. 

The \JWST SIDECAR software source code is unfortunately controlled under the International Traffic in Arms Regulations (ITAR) and subject to other restrictions.\footnote{The ITAR is a set of United States government regulations that pertain to specified defense-related technologies including \JWST's detector systems. Under ITAR, we cannot legally publish information that would facilitate duplicating the controlled technologies.} However, we are in the process of making the source code available to any United States Government Agency through the NASA Technology Transfer Program. Depending on the specific circumstances, release to other U.S. persons or organizations may be possible. To provide more insight into the clocking pattern, we have written an executable Jupyter notebook (python language) that is freely available for download on the \href{https://jwst.nasa.gov/publications.html}{\JWST web site}. Please contact the lead author for more information.

Building on \S~\ref{sec-clocking}, \S~\ref{sec-equations} presents the equations that are needed to reference-correct \IRSSquare data. Of these, the most important are Eqns.~\ref{eq-alphas2}--\ref{eq-betas2} and Eq.~\ref{eq-irs2cor}. These are the equations for determining the frequency dependent weights and applying them respectively. The other equations in this section are preliminaries to these.

Finally, we close with a summary. Readers who wish to understand \IRSSquare in detail are strongly encouraged to download the IDL source code and sample data. This ``hands-on'' information provides better insight into the details than any narrative can hope to achieve.

\section{Benefits and Downsides of \IRSSquare}\label{sec-benefits}

\begin{figure*}[t]
\begin{center}
\includegraphics[width=\textwidth]{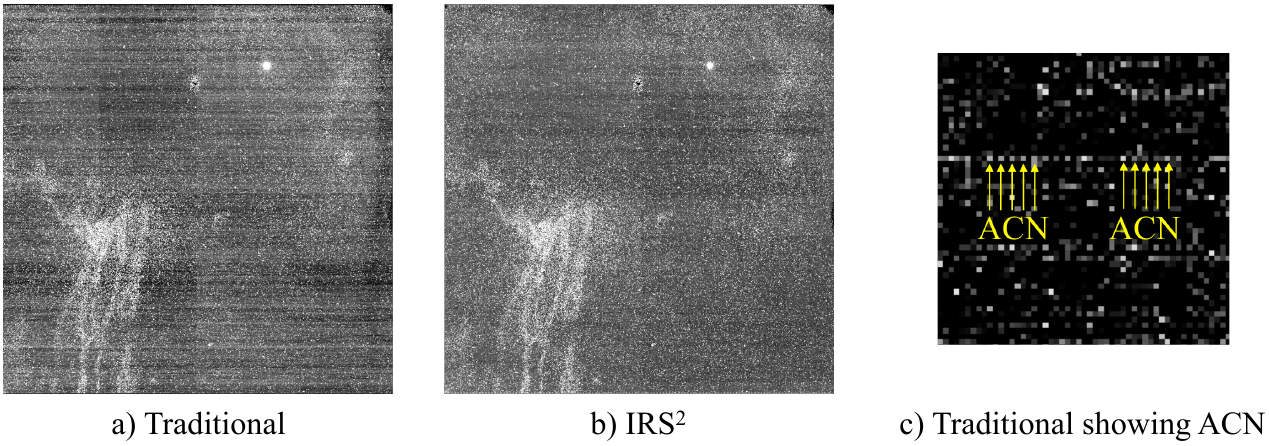}
\caption{This figure compares a) traditional and b) \IRSSquare readout for a pre-2010 engineering grade NIRSpec H2RG. The clouds of white ``warm pixels'' are the result of a detector degradation issue that was completely solved several years ago \citep{Rauscher:2012dc}. Both images are on the same $\left[-10~e^-,+20~e^- \right]$ grayscale. The correlated noise (faint horizontal banding) is clearly lower in the \IRSSquare image. Panel c) shows a subarray extracted from a traditional image showing ACN. \IRSSquare's interleaved reference pixels mitigate ACN by including even and odd reference pixel samples.}\label{fig-side-by-side}
\end{center}
\end{figure*}

For \JWST NIRSpec, the most important benefit of \IRSSquare is suppressing correlated noise. The noise appears primarily as horizontal banding oriented perpendicular to the spectral dispersion direction. This corresponds to the H2RG's fast scan direction.  The faint banding is caused by $1/f$ drifts in the SIDECAR ASIC readout electronics.  Another important correlated noise source is alternating column noise (ACN). The ACN is a Teledyne proprietary artifact of how the even and odd column buses are implemented in HxRG readout integrated circuits (ROIC). Figure~\ref{fig-side-by-side} shows some examples of how correlated noise appears in traditionally sampled and \IRSSquare sampled NIRSpec data.

Readers who are interested in understanding more about H2RG read noise in general may wish to see \citet{Rauscher:2015kt}. This paper also includes a freely downloadable python language noise generator that can be used to produce realistic read noise for traditionally sampled H2RG systems. 

As highlighted in the introduction, for NIRSpec the lower noise variance and reduced correlated noise translate directly into better scientific performance. For individual objects, \IRSSquare provides higher signal-to-noise per unit observing time. The lower correlated noise that \IRSSquare provides enables novel high multilplex observing strategies using non-local sky samples.

One might ask, ``what are the possible downsides of \IRSSquare?'' For NIRSpec, implementing \IRSSquare required somewhat more development time (with associated cost) on account of the added complexity. Regarding technical downsides, \IRSSquare sends more data to the ground. In the NIRSpec implementation, previously existing \JWST data volume constraints (in other parts of the system) limited us to 65 up-the-ramp frames compared to 88~up-the-ramp frames in traditional mode. However, the reduced number of up-the-ramp frames is not fundamental to \IRSSquare. This could be fixed in a new system that did not have the same constraints.

\section{Why \IRSSquare Works}\label{sec-concept}

Astronomical detector systems are designed to be very stable, both in their response to light and read noise. In other words, they are designed so that, to a very high degree of approximation, the read noise's covariance matrix is independent of time. The read noise is therefore very nearly covariance stationary. It is a classical result that the eigenspace of a stationary covariance matrix is Fourier space. The Fourier basis vectors provide an orthogonal representation of the read noise.

This is important because, when stationary noise is viewed in Fourier space, the different frequencies are uncorrelated. Even though the data may contain strong pixel-to-pixel correlations when viewed in the time domain (as images), there are no correlations between frequencies when the same data are viewed in Fourier space.

With this realization, we understood that we could treat each frequency independently to model the regular pixels as a linear combination of all available reference information. In the current implementation, the model includes the reference pixels (including additional reference samples) and the reference output. If other reference time series were to become available, we could easily modify the \IRSSquare equations to include them.


\section{\JWST \IRSSquare SIDECAR Assembly Code Implementation}\label{sec-clocking}

\subsection{Simple Prototype Implementation}\label{sec-clocking-proto}

\begin{figure*}[t]
\begin{center}
\includegraphics[width=0.8\textwidth]{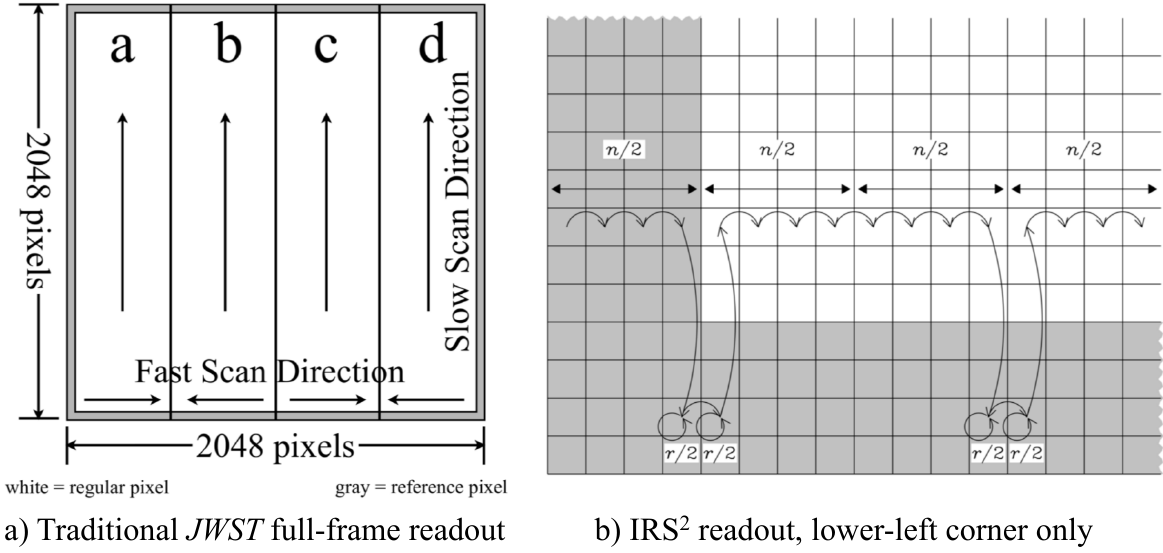}
\caption{a) In traditional H2RG clocking, pixels are read out and digitized in the same order as they appear physically on the H2RG ROIC. For traditional \JWST clocking, the H2RG is read out using four video outputs. These appear as thick vertical bands here. The fast scanners alternate directions and readout proceeds one row a time from the bottom to the top. The photo-sensitive area is bordered on all sides by a 4-pixel wide frame of reference pixels. In b), we zoom in on the lower left hand corner of the H2RG to show how the \IRSSquare clocking pattern differs to interleave reference and normal pixels. Every $n$ normal pixels, we step out to a reference row to digitize an even numbered reference pixel $r/2$ times and then an odd numbered reference pixel $r/2$ times. We sample both even and odd reference columns in order to subtract alternating column pattern noise at the Nyquist frequency \citep{Moseley:2010kc}. In \IRSSquare, the reference output is digitized and saved in parallel with the four video outputs. The H2RG's vertical window mode scanner is used to implement stepping out to the reference pixels, which is effectively no different from reading a 1-pixel subarray that happens to be in the same column as the science pixel.  This figure omits some timing overheads for clarity. Please see Appendix~\ref{sec-timing-detail} for more information on the timing, including an explicit discussion of the NIRSpec timing overheads.}\label{fig-clocking}
\end{center}
\end{figure*}

\begin{figure}[t]
\begin{center}
\includegraphics[width=\columnwidth]{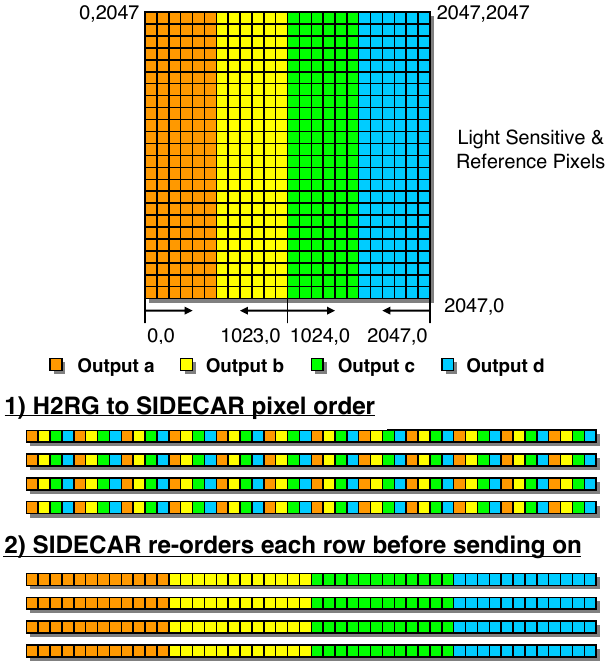}
\caption{This figure shows the traditional full frame H2RG readout in which the reference output is subtracted in real time in the SIDECAR's video pre-amplifiers. Hence, the reference output does not appear here. 1) Outputs a, b, c, and d are sampled and digitized simultaneously. 2) They are de-interlaced in the SIDECAR prior to being send on for further processing. This figure contains some of the same information as figure~\ref{fig-clocking}a.  We include it because it highlights how the data are transmitted in the flight data systems. This will serve as a useful point of comparison to figure~\ref{fig-pixel-order2}, which does the same for \IRSSquare readout.}\label{fig-pixel-order}
\end{center}
\end{figure}

\begin{figure*}[t]
\begin{center}
\includegraphics[width=\textwidth]{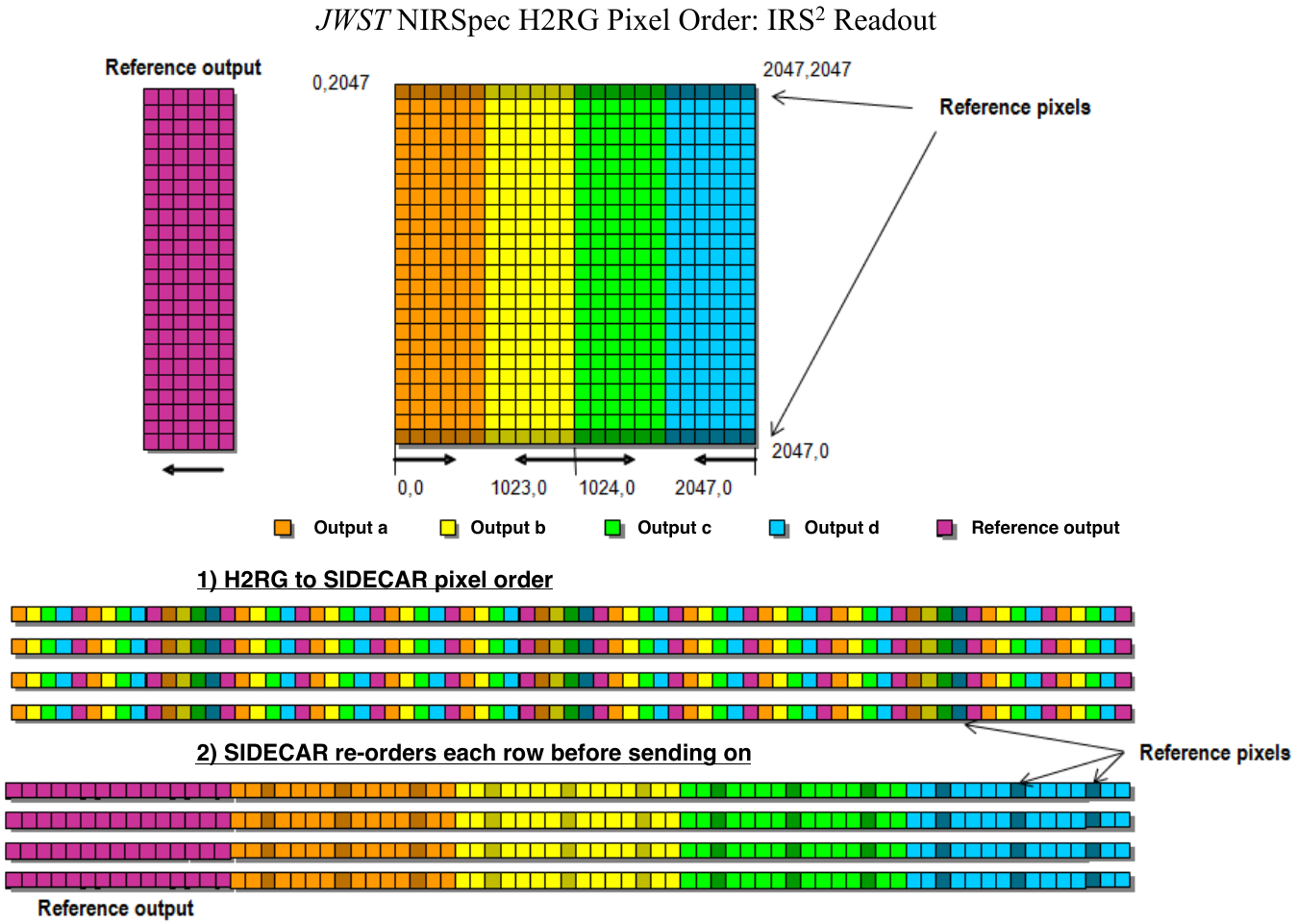}
\caption{This figure should be contrasted to figure~\ref{fig-pixel-order}. In \IRSSquare, shown here, the reference output is sampled and digitized in parallel with the regular outputs a, b, c, and d. These are de-interlaced in the SIDECAR as before prior to being send on for further processing. In this figure, dark shading indicates reference pixels. The H2RG has four rows of reference pixels on the top and bottom edges. Here we highlight all reference rows with dark shading. In practice, only one of the eight reference rows is selected and used to provide all interleaved reference pixels.}\label{fig-pixel-order2}
\end{center}
\end{figure*}

Compared to the traditional full frame readout pattern, the \IRSSquare clocking pattern contains two distinct differences. First is the interleaving of H2RG reference pixels (from one specified reference row) within the science data stream. This is shown in figure~\ref{fig-clocking}. Second is the separate digitization of the H2RG reference output (\ie ``single-ended readout'') to facilitate differential readout in ground post-processing instead of in the SIDECAR pre-amplifier block (figures~\ref{fig-pixel-order}-\ref{fig-pixel-order2}).

``Stepping out'' from the regular pixels to the reference row and ``stepping in'' again is accomplished using the H2RG's moveable guide window. In effect, we program a moveable 1-pixel guide window in a pre-selected reference row that shadows the regular pixels. The full-frame horizontal scanner is always used, while the full-frame vertical scanner is used for regular pixels and the guide window vertical scanner is used for the reference row.

We began developing \IRSSquare at a time when \JWST flight software development was already well underway. To implement this concept within the existing \JWST flight system, multiple software components needed to be enhanced including the SIDECAR assembly code and the \JWST Integrated Science Instrument Module's (ISIM) Command and Data Handling Software (ICDH).     Some of these enhancements would be applicable to any HxRG/SIDECAR system, and others are only necessary to work-around existing downstream hardware limitations on \JWST. 


The \JWST prototype and flight implementation allows for user configurable parameters to specify the details of the readout ($n$ = number of normal science pixels and $r$ = the number of interleaved reference pixels), and the row within the H2RG to use as reference pixels.  Figure~\ref{fig-clocking}b shows how $n$ and $r$ appear in the timing pattern. The reference row can be selected from the bottom or top four rows, all of which are reference rows. Through experimentation with both the prototype (engineering grade) and NIRSpec flight hardware, \IRSSquarePars{16}{4} was determined to be near optimal for NIRSpec. In practice, we found that it did not matter greatly which reference row was selected.

\begin{figure*}[t]
\begin{center}
\includegraphics[width=\textwidth]{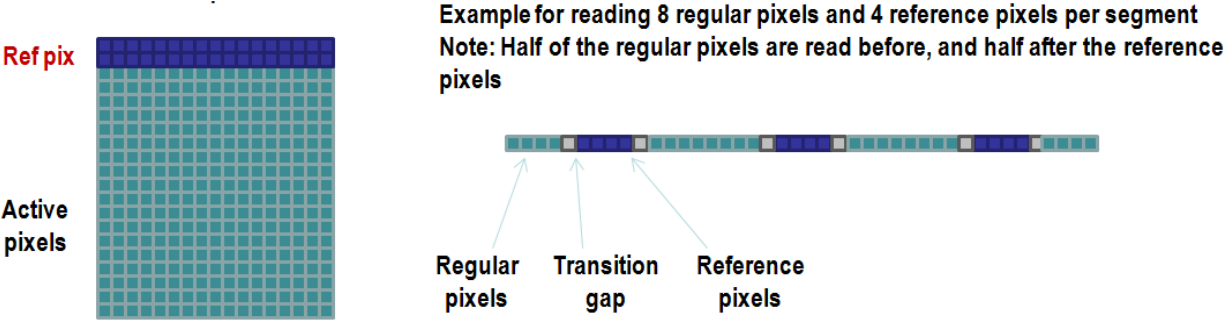}
\caption{There is a one pixel time ``transition gap'' associated with stepping out to the reference pixels or stepping back in to the science pixels.}\label{fig-transition-gap}
\end{center}
\end{figure*}

Implementation of the interleaved pixel readout on the \JWST hardware utilizes the H2RG's vertical window mode scanner, which was originally intended for support of guide mode.    Before the exposure begins, the SIDECAR assembly code positions the vertical window mode scanner at the user specified row within the H2RG as part of preparation of the acquisition.    During readout, the transition from the science pixels to the reference pixels is done by selecting the appropriate vertical scanner. The single full frame horizontal scanner is utilized for both science and reference pixel readouts. The time needed to transition between vertical scanners injects a single pixel time gap between the sampling of the science and reference pixels (figure~\ref{fig-transition-gap}).

Finally, to implement the single-ended digitization of the H2RG reference output, the signal is routed via the SIDECAR's global routing bus into a dedicated video channel in parallel with the other four video channels.   In contrast to traditional readout, the five video channels are configured for single-ended operation. 
  
This implementation contains all readout changes needed to implement \IRSSquare on standard SIDECAR~+~H2RG system as was done with the \JWST prototype code, which operates with a JADE2, and free of constraints from the other flight elements of \JWST. For most readers, the prototype code will be the most straightforward to adapt to their own systems.
  
 \subsection{Additional Challenges in Flight Implementation}\label{sec-clocking-flight}
 
As discussed, the \JWST flight implementation had further complexities which arose due to the challenges of implementing \IRSSquare in the already-built flight systems.  Creative use of the values within the SIDECAR science packet headers were required to have the data flow properly  to the downstream electronics.   Each row of the H2RG was split in half for compatibility with the NIRSpec focal plane electronics, and each frame was also split in half to accommodate the memory limitations within the data system memory. Finally, the pixel ordering within the flight code needed to be updated significantly for \IRSSquare, again to work properly with the downstream electronics design. If a reader chooses to review the \JWST flight code, then these additional complexities should be accounted for, but are unlikely to manifest in other systems.


\section{Reference Correcting IRS$^2$ Data}\label{sec-equations}

After acquiring the data, \IRSSquare reference correction should be the first step in the calibration process. Once the raw data cubes have been reference corrected, subsequent calibration steps including linearity correction and up-the-ramp slope fitting can be done using the standard tools that are available at most observatories.

For NIRSpec, we have found that it is not possible to improve upon \IRSSquare reference correction with any further reference pixel correction. As expected, subsequent use of the reference rows and columns degrades the noise by at least a few percent. If attempted, the degradation typically appears as a small increase in the correlated noise.

The H2RG provides three types of output; $n_p$ normal pixels, $r_p$ reference output, and $\rho_p$ (``rho sub p'') reference pixels. In \IRSSquare, these are represented by vectors where $p$ is a time domain index that runs over all time steps in the exposure (and incidentally over all pixels since they are time-ordered). In \citet{Moseley:2010kc}, we showed that the read noise's eigenspace is close to Fourier space. We therefore work in Fourier space because the basis vectors (sines and cosines) are linearly independent. Fourier transforming each output, we arrive at $n_\nu$, $r_\nu$, and $\rho_\nu$, where $\nu$ is is an index that runs over frequency.

IRS$^2$ is a linear model. The reference corrected normal pixels are represented by, \begin{equation}
n_\nu^\prime = n_\nu -\alpha_\nu r_\nu -\beta_\nu\rho_\nu,\label{eq-linmod}
\end{equation}
 where $\alpha_\nu$ and $\beta_\nu$ are vectors of frequency dependent weights that function analogously to Wiener filters.

If we have a sufficiently large number of training dark frames (typically $\geq 10^3$ for NIRSpec), and $i$ is an index that runs over this set, then we can use the method of least squares to solve for $\alpha_\nu$ and $\beta_\nu$. Let \begin{equation}
X_\nu^2 = \sum_i \left(n_\nu^i -\alpha_\nu r_\nu^i -\beta_\nu\rho_\nu^i\right)\left(n_\nu^{i*} -\alpha_\nu^* r_\nu^{i*} -\beta_\nu^*\rho_\nu^{i*}\right),\label{eq-fom}
\end{equation}
 be the least squares figure of merit, where $^*$ denotes the complex conjugate. $X_\nu^2$ is minimized when, \begin{eqnarray}
\alpha_{\nu}& =& \frac{\left(\sum_{i}n_{\nu}^{i}r_{\nu}^{i*}\right)\left(\sum_{i}\rho_{\nu}^{i}\rho_{\nu}^{i*}\right)-\left(\sum_{i}n_{\nu}^{i}\rho_{\nu}^{i*}\right)\left(\sum_{i}\rho_{\nu}^{i}r_{\nu}^{i*}\right)}{\left(\sum_{i}r_{\nu}^{i}r_{\nu}^{i*}\right)\left(\sum_{i}\rho_{\nu}^{i}\rho_{\nu}^{i*}\right)-\left(\sum_{i}\rho_{\nu}^{i}r_{\nu}^{i*}\right)\left(\sum_{i}r_{\nu}^{i}\rho_{\nu}^{i*}\right)}\label{eqn-alpha}\\
&\textrm{and}\nonumber\\
\beta_{\nu}& =& \frac{\left(\sum_{i}n_{\nu}^{i}\rho_{\nu}^{i*}\right)\left(\sum_{i}r_{\nu}^{i}r_{\nu}^{i*}\right)-\left(\sum_{i}n_{\nu}^{i}r_{\nu}^{i*}\right)\left(\sum_{i}r_{\nu}^{i}\rho_{\nu}^{i*}\right)}{\left(\sum_{i}r_{\nu}^{i}r_{\nu}^{i*}\right)\left(\sum_{i}\rho_{\nu}^{i}\rho_{\nu}^{i*}\right)-\left(\sum_{i}\rho_{\nu}^{i}r_{\nu}^{i*}\right)\left(\sum_{i}r_{\nu}^{i}\rho_{\nu}^{i*}\right)}.\label{eqn-beta}
\end{eqnarray}
\period

One can factor Eqns.~\ref{eqn-alpha}-\ref{eqn-beta} into a convenient set of sums that can be augmented whenever new darks become available to improve the coefficients.\footnote{If a substantial change is made to the system (\eg changing a bias voltage or the operating temperature), we recommend recalibrating the \IRSSquare coefficients using new training data.} These are as follows.
\begin{eqnarray}\label{eq-sums}
N_{\nu}& =& \sum_{i} n_{\nu}^{i}n_{\nu}^{i*}\\
R_{\nu}& =& \sum_{i} r_{\nu}^{i}r_{\nu}^{i*}\\
P_{\nu}& =& \sum_{i} \rho_{\nu}^{i}\rho_{\nu}^{i*}\\
X_{\nu}& =& \sum_{i} n_{\nu}^{i}\rho_{\nu}^{i*}\\
Y_{\nu}& =& \sum_{i} n_{\nu}^{i}r_{\nu}^{i*}\\
Z_{\nu}& =& \sum_{i} \rho_{\nu}^{i}r_{\nu}^{i*}
\end{eqnarray}
The vectors $N_\nu$, $R_\nu$, and $P_\nu$ are real while $X_\nu$, $Y_\nu$, and $Z_\nu$ are complex. 

With these definitions, we can rewrite Eqns.~\ref{eqn-alpha}-\ref{eqn-beta} as
\begin{eqnarray}
\alpha_\nu& =& \frac{Y P - X Z}{R P - Z Z^*},\label{eq-alphas}~\textrm{and}\\
\beta_\nu& =& \frac{X R - Y Z^*}{R P - Z Z^*}\label{eq-betas},
\end{eqnarray} where we have suppressed the $\nu$ suffix on the right hand side to achieve a more compact notation. In Eqns.~\ref{eq-alphas}-\ref{eq-betas}, the denominators are real but the numerators are in general complex. In general, $\alpha_\nu$ and $\beta_\nu$ are therefore complex.

\begin{figure*}[t]
\begin{center}
\includegraphics[width=\textwidth]{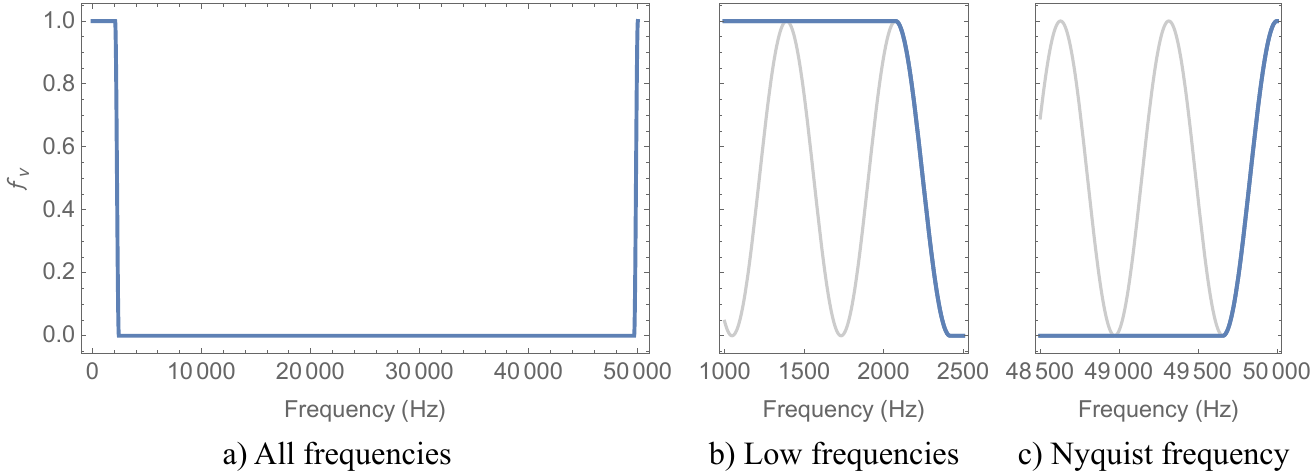}
\caption{The interleaved reference pixels, $\rho_\nu$, sample only low frequency noise and alternating column noise at the Nyquist frequency. To avoid injecting mid-frequency noise, we apply an apodized filter, $f_\nu$. Panel a) shows the filter that is used with the  standard NIRSpec $\left(n=16,r=4\right)$ \IRSSquare clocking pattern. b) We use one quadrant of a sine function to roll the response from unity to zero with the half power point at $f_{1/2}=2,247.19\rm~Hz$. The gray curve is a graphical element to highlight the sinusoidal shape of the roll off. c) This is mirrored at the Nyquist frequency to enable \IRSSquare removal of alternating column pattern noise. The width of the roll off and roll on is somewhat arbitrary at present. We empirically determined that 5,000 frequency steps $\left(=342.894\rm~Hz\right)$ works well for NIRSpec. Appendix B explains more fully how we determined the half power point for this particular \IRSSquare pattern.}\label{fig-filter}
\end{center}
\end{figure*}

At intermediate frequencies,  the interleaved reference pixels vector $\rho_\nu$, is not sampled. We therefore tailor $\beta_\nu$ with an apodized filter, $f_\nu$. Figure~\ref{fig-filter} shows the filter that we used while developing the standard NIRSpec \IRSSquare \IRSSquarePars{16}{4} clocking pattern. Appendix~\ref{sec-timing-detail} provides more information on how the filter was designed, and specifically on how the roll off frequency was chosen.

Including the filter, Eqns.~\ref{eq-alphas}--\ref{eq-betas} can be rewritten \begin{eqnarray}
\alpha_\nu& = & \frac{Y - f f^* X Z/P}{R - f f^* Z Z^*/P},~\textrm{and}\label{eq-alphas2}\\
\beta_\nu& = & f^* \frac{X R/P - Y Z^*/P}{R - f f^* Z Z^*/P}.\label{eq-betas2}
\end{eqnarray}\period Once the coefficients have been determined using Eqns.~\ref{eq-alphas2} and \ref{eq-betas2}, Eq.~\ref{eq-linmod} simplifies to, \begin{equation}
n_{p}^\prime = n_{p} - \mathcal{F}^{-1}\left[\alpha \mathcal{F}\left(r_{p}\right)+\beta \mathcal{F}\left(\rho _{p}\right)\right]\label{eq-irs2cor},
\end{equation}
were $\mathcal{F}$ is the familiar discrete Fourier Transform. There is no need to Fourier transform the normal pixels in the pipeline. Moreover, one could implement these operations as convolutions in the time domain if desired.

This concludes the core set of equations that are needed to implement IRS$^2$. The frequency dependent weights are inferred from a set of training darks using Eqns.~\ref{eq-alphas2} and \ref{eq-betas2}. Once these weights are known, Eq.~\ref{eq-irs2cor} is used to apply them to the data.

\subsection{Understanding $\alpha_\nu$ and $\beta_\nu$}\label{sec-alphas-betas}

\begin{figure*}[t]
\begin{center}
\includegraphics[width=\textwidth]{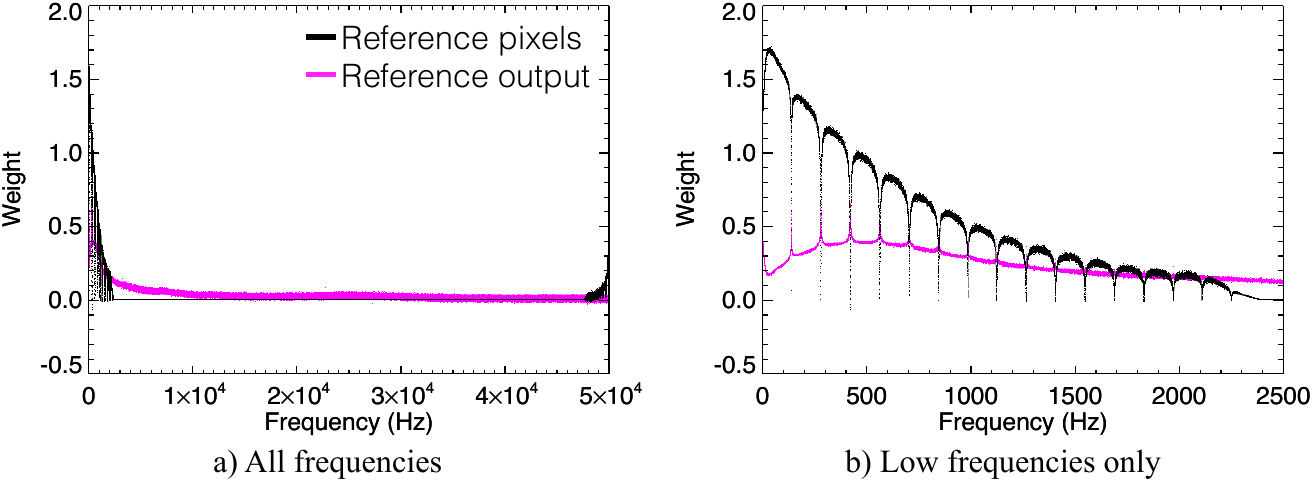}
\caption{In \IRSSquare, the reference output and interleaved reference pixels are filtered with a set of frequency dependent weights. Panel a) plots all frequencies for a non-flight prototype of the NIRSpec detector subsystem. It shows that the reference output, $\alpha_\nu$, is important at all frequencies, and largest at low frequencies. The interleaved reference pixels, $\beta_\nu$, have power only at very low frequencies and near the Nyquist frequency. There is a wing on the low frequency side of 50~kHz. This is caused by column switching acting as a carrier for much lower frequency $1/f$. Panel b) highlights the low frequencies. Interestingly, the reference pixels have gain $<1$ at all frequencies for this system. This partially explains why traditional \JWST readout, that subtracts the reference output with unity gain, is not optimal.}\label{fig-alphas-betas}
\end{center}
\end{figure*}

Recall that $\beta_\nu$ is a set of frequency dependent weights for the interleaved reference pixels and $\alpha_\nu$ is a similar vector for the reference output. Figure~\ref{fig-alphas-betas} shows the measured values for a prototype implementation of \IRSSquare (\ie not the flight system).

Figure~\ref{fig-alphas-betas}a plots all frequencies and shows that the interleaved reference pixels primarily correct very low frequencies, including $1/f$, and ACN at 50~kHz. The low frequency wing that is visible near 50~kHz  is caused by column switching acting as a carrier for much lower frequency $1/f$. The amplitude of $\beta_\nu$ is zero at intermediate frequencies, as expected given that these frequencies have been filtered out.

The reference output, $\alpha_\nu$, has amplitude at all frequencies, but it too is strongest at low frequency.  Figure~\ref{fig-alphas-betas}b highlights the low frequencies. This figure shows one of the more important early findings in \IRSSquare development. The reference output does not have unity gain at low frequency. The behavior shown here is fairly typical for NIRSpec H2RGs. The reference output should be subtracted with less than unity gain. In traditional \JWST readout it is subtracted with unity gain and this explains why it is not more effective at subtracting out $1/f$ noise in traditional readout.

\subsection{Optimal Use of the Reference Output when Interleaved Reference Pixels are not Available}\label{sec-refout-only}

For \JWST, we were in the fortunate position of being able to write our own SIDECAR ASIC software. This allowed us to implement the interleaved reference pixels that are a hallmark of \IRSSquare. For groups who are not in a position to do this, the \IRSSquare study nevertheless provided insight for how to best use the reference output.

\begin{figure*}[t]
\begin{center}
\includegraphics[width=\textwidth]{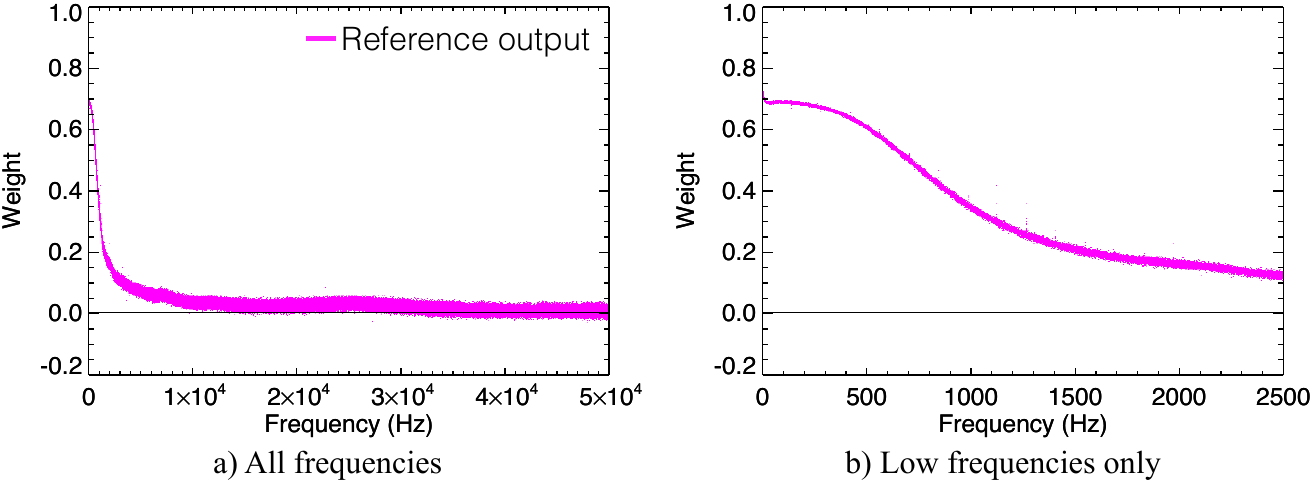}
\caption{If only the reference output is used to make the correction, then $\alpha_\nu$ needs to be recomputed. Here we show the result of recalculating $\alpha_\nu$ using the same training data as were used to produce figure~\ref{fig-alphas-betas}. The weight is less that unity at low frequencies. As expected, there is no weight at 50~kHz because the reference output does not experience column switching. Panel b) suggests a potentially simple way of improving the performance of existing systems by filtering the reference output and operating differentially. This is described more fully in the text.}\label{fig-alpha-only}
\end{center}
\end{figure*}

If only the  reference output is used to read an H2RG differentially, then $\alpha_\nu$ needs to be recalculated. Under these conditions, $\alpha_\nu$ takes all of the weight and is much closer to 1.0 at low frequencies (typically within ~20\%), although $\alpha_\nu$ still falls off at higher frequencies. Moreover, it contains no trace of any correction for the ACN since the reference output does not see the even and odd columns. Figure~\ref{fig-alpha-only} shows an example of recalculating $\alpha_\nu$ only using the same input data as were used in figure~\ref{fig-alphas-betas}. 

The considerable change at low frequency is because $r_\nu$ and $\rho_\nu$ are highly correlated at low frequencies, and thus either can be used to remove $1/f$ noise. When only the reference output is available, it takes all of the weight and therefore $\alpha_\nu$ approaches unity. The results shown in Figure~\ref{fig-alphas-betas} indicate that for this particular array, the interleaved reference pixels happened to be more effective at removing the lowest frequency noise.

We note in passing that figure~\ref{fig-alpha-only}b suggests that some HxRG systems could potentially be improved incrementally by applying a simple passive low pass filter to the reference output and operating the detector differentially. Although we have not experimentally tested this concept yet, we plan to do so in the near future. We do not expect filtering the reference output alone to be as powerful as \IRSSquare (for example it does nothing about ACN), but it may provide a simple way to boost the performance of existing systems without writing new controller software.

\subsection{Other Implementation Details}\label{sec-implementation-details}

Because IRS$^2$ is a linear model, a full implementation requires interpolating over all gaps and other non-linear events in the time series before applying these equations (\eg overheads at the ends of rows and frames as well as ``bad pixels'' and cosmic ray hits). In early prototypes, we used simple linear interpolation for this. We now use the somewhat more sophisticated interpolation scheme that can be seen in the source code. The interested reader is referred to the IDL source code that is available at \href{http://jwst.nasa.gov/publications.html}{http://jwst.nasa.gov/publications.html} for these implementation details.

For completeness, we note that if additional reference information that tracks the stationary read noise were to become available, then \IRSSquare could straightforwardly be extended to incorporate it. In this case, one would revise Eq.~\ref{eq-linmod} to include the new reference vector and update the other equations accordingly.

\section{Conclusion}\label{sec-summary}

Improved Reference Sampling and Subtraction (\IRSSquare; pronounced ``IRS-square'') is a technique for reducing the correlated read noise of near-IR detector systems. \IRSSquare was conceived, implemented, and tested by the \JWST Project at NASA Goddard for NIRSpec, which uses Teledyne H2RGs and SIDECAR ASICs. Compared to ``traditional'' HxRG readout, \IRSSquare uses a new clocking pattern to interleave many more reference pixels into the data than is otherwise possible. As part of the \IRSSquare post processing, \IRSSquare subtracts both the reference pixels and reference output using a set of least squares optimized frequency dependent weights. These weights were measured for the as built hardware using the equations of $\S$~\ref{sec-equations} to least squares fit the references to an extensive training data set.

For NIRSpec, \IRSSquare's primary benefit is to significantly reduce correlated noise. Compared to traditional \JWST readout, \IRSSquare images are cosmetically cleaner, with fewer instrument signatures (less $1/f$ banding, less alternating column noise, \etc). The cosmetically cleaner images will allow the use of more distant sky samples, thereby increasing MOS multiplex advantage in crowded fields. The cosmetically cleaner images should likewise increase the efficiency of integral field unit (IFU) observations by increasing the signal to noise ratio for realistic sky subtraction scenarios of extended sources. Our simulations suggest that SNR gains as great as 45\% per unit observing time are potentially possible depending on the source. In any case, for a read noise limited instrument like NIRSpec, lower noise is always better. We recommend that readers who are interested in exploring noise trades download the sample data because the results depend critically on the observing scenario. 

As an aid to groups who may wish to explore the benefits of \IRSSquare, we are making our prototype \IRSSquare calibration software and sample \JWST NIRSpec data freely available for download. These can be found on the \href{https://jwst.nasa.gov/publications.html}{\JWST web site}. Although the SIDECAR ASIC  detector readout software are ITAR sensitive (and subject to other controls), they are nevertheless available to other United States Government Agencies. Depending on the circumstances, we may be able to release the SIDECAR software to other United States entities. Please contact the lead author for more information.

Looking to the future, Teledyne coordinated with us while developing the H4RG series of near-IR detector arrays. The H4RGs build-in one \IRSSquare readout pattern. Teledyne refers to this as ``interleaved reference pixel readout''. If the new mode works as hoped, this would free users from having to write the kind of \IRSSquare clocking software that is described in $\S$~\ref{sec-clocking}. We plan to explore the new readout mode using the techniques and equations that are described in $\S$~\ref{sec-equations} in the near future. 

Looking further into the future still, we are eager to explore the utility of blanked off regular pixels in NIRSpec scenes by extending \IRSSquare to include them. In principle, blanked off regular pixels are an even better proxy of regular pixels than reference pixels. We plan to do this using a similar least squares approach once appropriate NIRSpec science data start to become available in mid-late 2019.

\acknowledgments
This work was supported by NASA as part of the James Webb Space Telescope Project. The authors wish to thank Drs. Pierre Ferruit and Marco Sirianni of the European Space Agency for carefully reading the entire manuscript and making several very helpful suggestions. We also wish to thank the referee for several useful comments, helpful ideas for providing more insight into the SIDECAR clocking software within the constraints of the ITAR.

\bibliographystyle{yahapj}
\bibliography{references.bib}

\appendix
\section{\JWST NIRSpec \IRSSquare Data Format}

This appendix describes the prototype data format that was used to generate the downloadable sample data. The flight format may differ. Please consult the NIRSpec instrument documentation at the Space Telescope Science Institute (STScI) for information on the flight data format.

The traditional \JWST clocking pattern reads the H2RG detectors using four outputs. The resulting $2048\times 2048$~pixel ``frames'' of data appear in thick, $2048\times 512$~pixel stripes (figure~\ref{fig-clocking}). Because \IRSSquare returns the reference output and the four regular outputs with additional reference pixels interleaved, the resulting frame format is different.

\begin{figure*}[t]
\begin{center}
\includegraphics[width=0.6\textwidth]{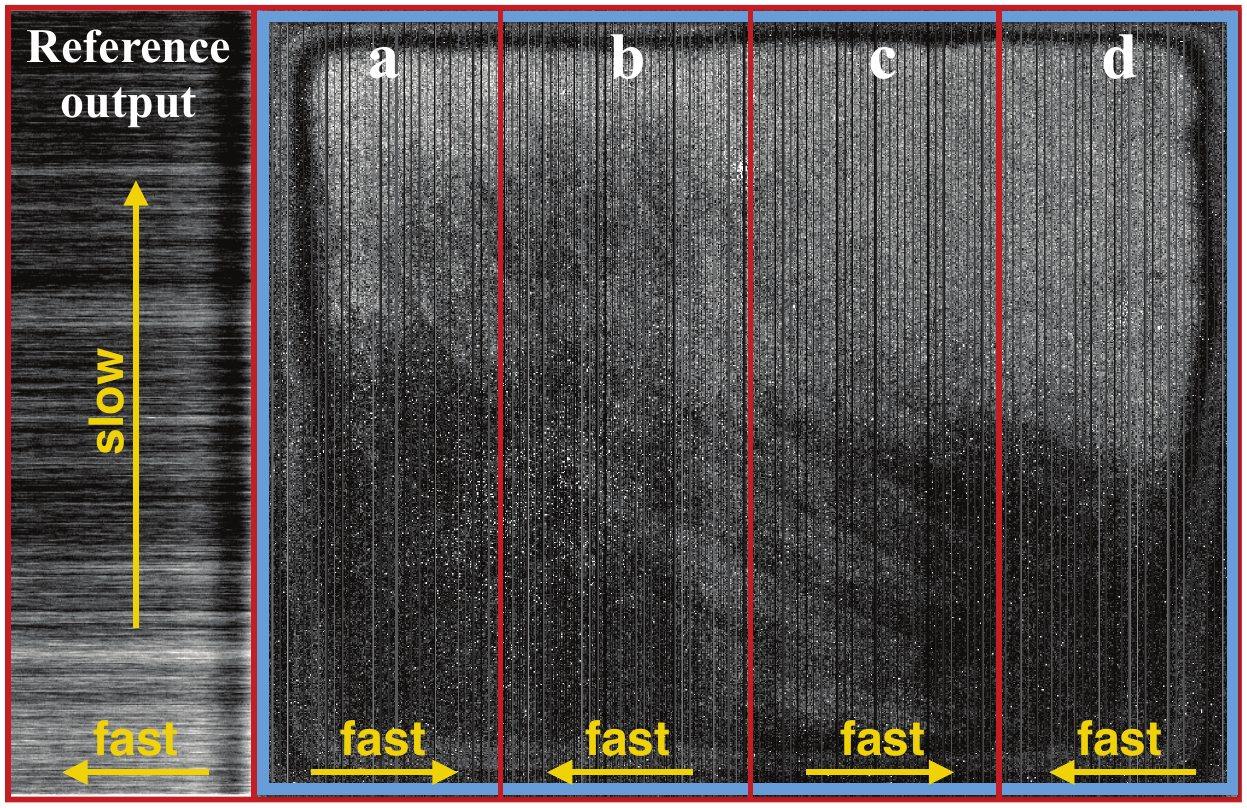}
\caption{This figure shows the \JWST NIRSpec \IRSSquare data format for one frame of H2RG data. To more clearly show fine structure, we have zero offset and renormalized each output. The fast and slow scanner directions are shown in yellow. The usual four pixel wide border of reference pixels that surrounds each frame of H2RG data is shown in blue. Data from the reference output appears at left. The four regular outputs are labeled a, b, c, and d. Each output is wider than 512~pixels wide because it includes regular pixels and the extra interleaved reference pixels. As in traditional readout, each output is 2048 pixels high. If one looks carefully at any of the outputs a-d, there are faint vertical stripes where the interleaved reference pixels appear at a lower signal level than the regular pixels on the same output.}\label{fig-data-format}
\end{center}
\end{figure*}

Figure~\ref{fig-data-format} shows one frame of \IRSSquare sampled data. In the default $\left(n=16,r=4 \right)$ \IRSSquare configuration, the resulting frame size is $2048\times 3200$~pixels.

\section{More Information on Pixel Timing and Filtering Interleaved Reference Pixels}\label{sec-timing-detail}

\begin{figure}[t]
\begin{center}
\includegraphics[width=.6\columnwidth]{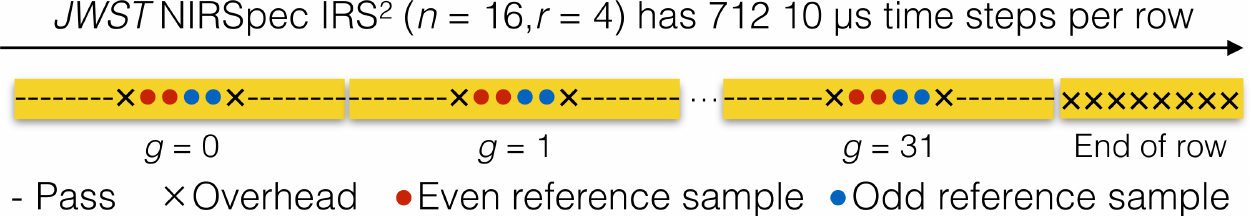}
\caption{This figure shows a timeline of the standard NIRSpec $\left(n=16,r=4\right)$ \IRSSquare clocking pattern for one row of one video output. There are twenty two $10~\mu$s time steps within each \IRSSquare group, $g$. When sampling regular pixels, the reference vector ``passes'' over them leaving gaps in the reference pixel time series. Other gaps include a one pixel time overhead for stepping out from the regular pixels to the reference pixels, and another one pixel overhead when stepping back in to the regular pixels. At the end of each row, there is an eight pixel time overhead for starting the next row. The corresponding row rate is 140.449~Hz. We selected this pattern to sample (mostly) $1/f$ noise at frequencies $\lesssim 3$~kHz and alternating column pattern noise at the 50~kHz Nyquist frequency. We were also required to select values of $n$ and $r$ that were practical to implement in the already built \JWST data systems.}\label{fig-refrow-timing}
\end{center}
\end{figure}

\JWST's H2RG detectors use a 100~kHz pixel clock, resulting in a 50~kHz Nyquist frequency. Early \IRSSquare studies \citep{Moseley:2010kc,Rauscher:2011jx} showed that the reference pixels were correlated with the regular pixels for frequencies lower than (very roughly) 3~kHz, and again near 50~kHz. At intermediate frequencies, the reference pixels do not correlate well with the regular pixels. Informed by this, and based on practical considerations for implementing \IRSSquare in the already built \JWST data systems, we selected \IRSSquarePars{16}{4} as the standard NIRSpec \IRSSquare clocking pattern. Figure~\ref{fig-refrow-timing} presents the timing sequence for one row of one output on a timeline.

In the standard \IRSSquarePars{16}{4} \IRSSquare pattern, $r=4$ reference samples are interleaved for every $n=16$ normal pixels. The pattern acquires two samples of a reference pixel in an even numbered column and two samples of the next reference pixel, which is in an odd numbered column. The ``even'' and ``odd'' reference  samples are taken to remove ACN. ACN originates in a well understood, but Teledyne proprietary aspect of how the H2RG columns are biased in the readout integrated circuit (ROIC).

This pattern leaves gaps in the reference pixel time sequence that must be interpolated over. Simple linear interpolation will work, although we are now using the more sophisticated interpolation scheme that can be seen in the downloadable IDL code. The gaps include time for ``passing'' over the regular pixels, and also one pixel overheads for ``stepping out'' from the regular pixels to the reference pixel and ``stepping in'' again. At the end of each row, there is an eight pixel time overhead for starting the next row.  After accounting for gaps, in each 22~pixel group, only the four reference pixel samples are saved, the first two in an even column and the second two in the next reference column.

Because the reference pixels do not correlate with the normal pixels at intermediate frequencies, \IRSSquare filters out the intermediate frequencies to avoid adding noise. Figure~\ref{fig-filter} shows the filter. Here we provide more detail on why the half power frequency was set to $f_{1/2}=2,247.19$~Hz for \IRSSquarePars{16}{4}. This represents the (loosely speaking) ``Nyquist'' frequency of the sampling pattern including the effects of gaps.

\begin{figure*}[t]
\begin{center}
\includegraphics[width=0.8\textwidth]{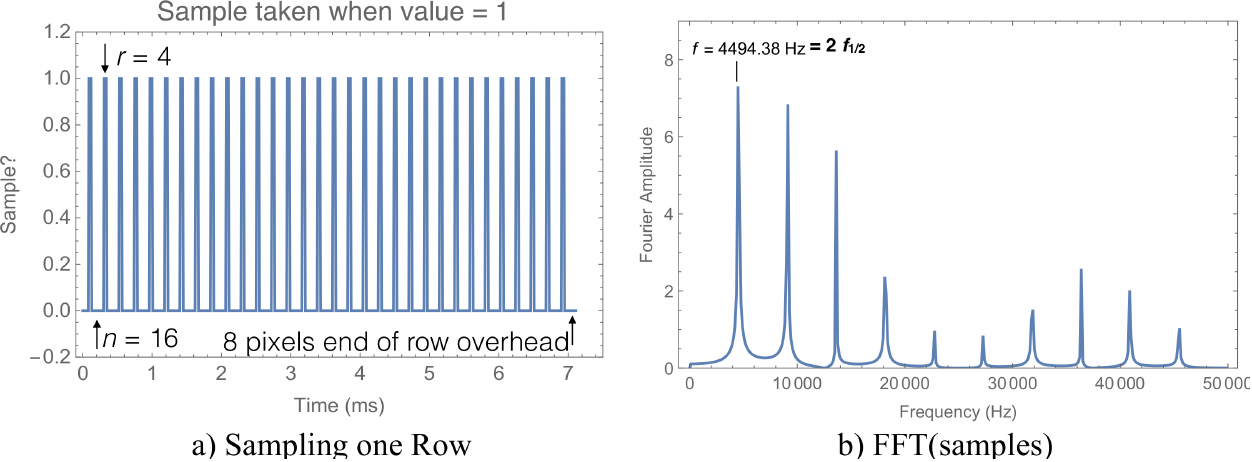}
\caption{In NIRSpec \IRSSquare $\left(n=16,r=4\right)$ readout at the 100~kHz pixel rate, the sampling interval for the interleaved reference pixels is $22\times \tau_{\rm pixel}=220~\mu$s. The corresponding sampling frequency is 4,545.45~Hz. However, a) because of the 8~pixel end of row overhead, the samples are  not equally spaced in time. To compute an effective ``Nyquist'' frequency for the unequal sampling, we b) compute the FFT of the sampling function (panel a) and measure the location of the first maximum. The half power point of the filter shown in Figure~\ref{fig-filter} is set to one half this frequency, $f_{1/2}=2,247.19$~Hz. In practice, this is only slightly different than the more usual ``half the sampling frequency'' rule, which would have worked about as well.}\label{fig-half-power-plot}
\end{center}
\end{figure*}

Figure~\ref{fig-half-power-plot} explains how the half power frequency was calculated. In panel a), we define a reference pixels vector that includes all time steps within one row. The value was set $=0$ when ``passing'' over a normal pixel or clocking overheads. For the four reference pixel samples in each group, the value was set $=1$. Panel b) shows the FFT of this vector. The maximum value occurs at $2f_{1/2}=$~Hz.

\end{document}